\documentclass[aps,prb,reprint]{revtex4-1}
\usepackage{graphicx}

\begin{document}


\title{Thickness-dependent phase transition in graphite under high magnetic field}


\author{Toshihiro Taen}
\email[]{taen@issp.u-tokyo.ac.jp}
\affiliation{The Institute for Solid State Physics, The University of Tokyo, Kashiwa, Chiba, 277-8581}
\author{Kazuhito Uchida}
\affiliation{The Institute for Solid State Physics, The University of Tokyo, Kashiwa, Chiba, 277-8581}
\author{Toshihito Osada}
\affiliation{The Institute for Solid State Physics, The University of Tokyo, Kashiwa, Chiba, 277-8581}


\date{\today}

\begin{abstract}
Various electronic phases emerge when applying high magnetic fields in graphite.
However, the origin of a semimetal-insulator transition at $B \simeq 30$ T is still not clear,
while an exotic density-wave state is theoretically proposed.
In order to identify the electronic state of the insulator phase,
we investigate the phase transition in thin-film graphite samples that were fabricated on silicon substrate by a mechanical exfoliation method.
The critical magnetic fields of the semimetal-insulator transition in thin-film graphite shift to higher magnetic fields,
accompanied by a reduction in temperature dependence.
These results can be qualitatively reproduced by a density-wave model by introducing a quantum size effect. 
Our findings establish the electronic state of the insulator phase as a density-wave state standing along the out-of-plane direction,
and help determine the electronic states in other high-magnetic-field phases.
\end{abstract}


\maketitle

\section{INTRODUCTION}
\begin{figure}
\includegraphics[width=8cm]{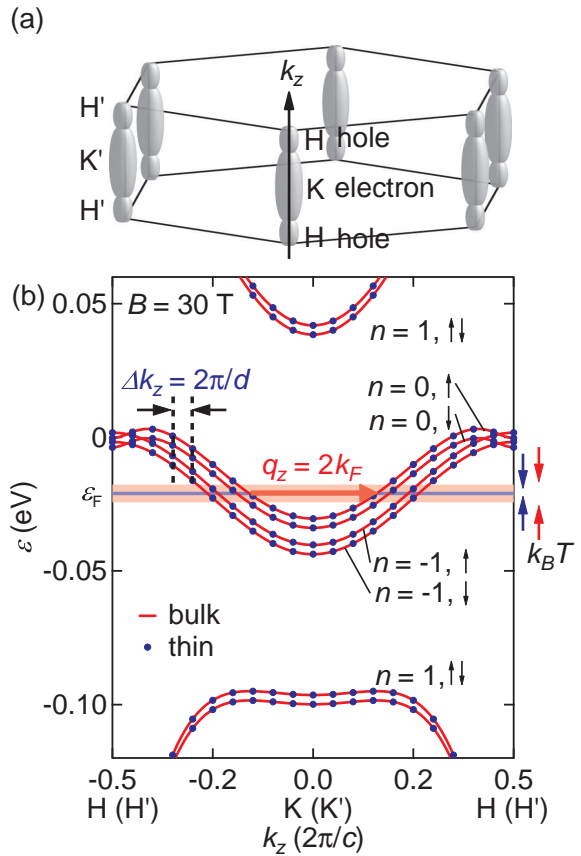}
\caption{(a) Schematic view of Brillouin zone and Fermi surfaces at $B=0$ in graphite. Electron pockets and hole pockets are formed around K (K$^{\prime}$) and H (H$^{\prime}$) points, respectively. Valleys along H-K-H and H$^{\prime}$-K$^{\prime}$-H$^{\prime}$ lines are energetically degenerated. The size of Fermi pockets is exaggerated for clarity. (b) Calculated Landau subbands in graphite under magnetic field $B = 30$ T along the $c$-axis. Only four Landau subbands are at the Fermi level $\varepsilon_F$. Calculation is based on the Slonczewski-Weiss-McClure model with $\gamma_3 = 0$ (Refs.~\onlinecite{PhysRev.109.272,PhysRev.119.606,JPhysSocJpn.17.808}). Red lines indicate the dispersions for the bulk (thick enough) system, and blue markers are for the thin-film system. Width of horizontal light blue and light red bars indicate $k_B T$ at low and high temperatures, respectively (see main text). The density-wave state characterized by $q_z=2k_F$ is expected to appear in the bulk system, while it is expected to be unstable in the thin-film system owing to the sparse $k_z$ states.}
\label{f1}
\end{figure}

A semimetal-insulator transition in graphite, discovered in the early 1980s (Ref.~[\onlinecite{Tanuma1981}]),
has recently regained much attention~\cite{PhysRevLett.110.266601,JPhysSocJpn.84.054709,PhysRevLett.103.116802,PhysRevLett.119.136601,SciRep.7.1733,CamargoEscoffier2018}.
This transition is induced by high magnetic fields of $B \simeq 30$ T along the $c$-axis (out-of-plane direction) at low temperatures.
Reflecting a low carrier density,
only four quasi-one-dimensional Landau subbands [$(n=0,\uparrow), (n=0,\downarrow), (n=-1,\uparrow)$, and $(n=-1, \downarrow)$] remain at the Fermi level under high magnetic fields~\cite{PhysRev.109.272,PhysRev.119.606,JPhysSocJpn.17.808} [the so-called quasiquantum limit; see red lines in Fig.\ref{f1}(b)],
which should be responsible for this electronic phase transition.
Taking the quasi-one-dimensionality and the electron-electron interaction into account,
Yoshioka and Fukuyama proposed the exotic density-wave state, \textit{valley-density wave} (VDW) state~\cite{JPhysSocJpn.50.725}, as illustrated in the following.
Graphite has two energetically equivalent band dispersions (so-called valleys) along H-K-H and H$^{\prime}$-K$^{\prime}$-H$^{\prime}$ lines in the reciprocal lattice ($k$) space,
which form an electron Fermi pocket around the K (K$^{\prime}$) point, and a hole Fermi pocket around H (H$^{\prime}$) point,
as can be seen in Fig.~\ref{f1}(a).
If we focus on one valley (e.g., H-K-H),
it forms a $2k_F$-type charge-density wave (CDW) along the $c$ axis direction under high magnetic fields along the $c$-axis.
In the counter part of the valley (e.g., H$^{\prime}$-K$^{\prime}$-H$^{\prime}$),
it also forms a CDW but is antiphase to the counter valley.
This means that, in total, the VDW has no spatial modulation of carrier to cancel out the direct Coulomb repulsive interaction, 
which is analogous to the spin-density wave (SDW) if we read the spin degrees of freedom as the valley ones.
Although there are some differences in detail,
all subsequent theories support the formation of the density-wave state~\cite{PhysicaB.201.384,PhysRevB.29.6722,JPhysCondensMatter.10.11315}.


On the other hand,
experimental verification of the density-wave state was a challenging problem.
First, if the ordered state is VDW,
it is impossible to directly observe it by utilizing, for example, x rays, since the spatial charge modulation should be absent or negligibly small.
Another common way of investigating the density-wave state is to detect non-Ohmic transport.
The nonlinearity was actually found in the in-plane~\cite{PhysRevLett.54.1182} and out-of-plane transport~\cite{JPhysSocJpn.68.181},
but its broad transition from a low conducting state to a high conducting state was ambiguous evidence for the sliding motion of the density wave.
In addition, it is not clear how to understand the in-plane transport results in the scenario of the density-wave standing along the out-of-plane direction.
It is noteworthy that swift neutron irradiation in graphite crystal was successful in controlling the phase boundary~\cite{JPhysSocJpn.68.1300,PhysicaB.298.546,JPhysConfSer.150.022099,JPhysCondensMattter.21.344207}.
In those experiments,
the transition line around $B\simeq 30$ T in the phase diagram shifted to higher magnetic fields almost in a parallel manner with the introduction of disorders.
This trend is basically understood by applying the theory of the ``pair-breaking effect,'' which is well-known in superconductivity~\cite{SolidStateCommun.52.975}.
This agreement manifests that some kind of pairing state is involved in the transition,
whereas it is difficult to provide a comprehensive interpretation of the formation of the density-wave state owing to concomitant carrier doping.

Recent discovery of a new electronic phase above $B > 53$ T offers a more confusing problem~\cite{PhysRevLett.110.266601}.
According to the Slonczewski-Weiss-McClure model,
which is known to accurately reproduce the band structure deduced from the quantum oscillations~\cite{PhysRevLett.102.166403},
the ($n=0, \uparrow$) subband escapes from the Fermi level at $B=53$ T (Refs.~[\onlinecite{PhysicaB.256.621,PhysicaB.201.384,JPhysCondensMattter.21.344207}]).
Therefore, it was believed that the anomalous electronic state will exist only between $B \simeq 30$ T and 53 T,
as the $(n=0,\uparrow)$ Landau subband is believed to be responsible for the density-wave formation in Ref.~[\onlinecite{JPhysSocJpn.50.725}].
In fact, the behavior of the in-plane resistivity seems to reenter the normal metallic state at 53 T (Ref.~[\onlinecite{PhysicaB.256.621}]).
However, according to Fauqu\'{e} \textit{et al.},
another high-resistivity state was found above 53 T by a longitudinal transport measurement ($R_{zz}$),
and the reentry of the conducting state needs to be as large as $B = 75$ T (Ref.~[\onlinecite{PhysRevLett.110.266601}]).
The authors proposed a new sequence of the Landau subband detachment to qualitatively explain the phenomena,
but there is no clear consensus as to that scenario so far~\cite{JPhysSocJpn.84.054709,PhysRevLett.119.136601}.
Even if other subbands are responsible for the phase transition,
a reasonable explanation for the anisotropic conducting state in the new phase,
and the reason for the location of the endpoint at $B=75$ T, are absent.

To unveil the true evolution of the electronic state under a high magnetic field,
it is significant to prove the electronic state as the density-wave state between $B \simeq 30$ T and 53 T.
In this study, we investigate the thickness dependence of the semimetal-insulator transition at $B \simeq 30$ T.
According to basic solid-state physics, 
the interval of $k$ points along the $z$-direction $k_z$ ($z||c$), $\Delta k_z$, is written as $\Delta k_z = 2\pi/d$, where $d$ is the thickness of the system.
If $d$ is sufficiently large compared with the lattice constant $c$,
the dispersion can be regarded as continuous [red lines in Fig.~\ref{f1}(b)].
With a reduction of thickness $d$,
the dispersion is no longer continuous owing to the quantum size effect,
as shown by blue markers in Fig.~\ref{f1}(b).
If some nesting vector of $q_z$ (the vector connecting $k_z$ points) is responsible for the phase transition in bulk graphite,
the formation of the density-wave state tends to be inhibited in the thin-enough sample owing to the sparseness of the $k_z$ points.
In fact, neither mono- nor bilayer system (graphene) shows an insulator transition in high magnetic fields.
Suppose that
a band-width of a few tens of millielectronvolts in the Landau subband is divided into a hundred points,
and the energy spacing is a few Kelvin,
which is comparable to the phase transition temperature.
Therefore, this level spacing effect is expected to appear on the order of hundreds of unit-cell-thick systems (roughly 70 nm).
We note that, in contrast to the neutron irradiation experiment,
this method is expected not to introduce additional disorders or carriers,
which is an advantage of the simple interpretation.
In fact, we confirmed it in our 80-nm-thick film through the evaluation of the residual resistivity ratio (RRR) and Dingle temperature ($T_D$).
The higher RRR and the smaller $T_D$ indicate high purity of samples.
The observed values were RRR $\simeq 6$ and $T_D = 3-7$ K, respectively.
These values are reasonably in good agreement with those in bulk samples (RRR $>10$ and $T_D = 0.5-4$ K, respectively~\cite{Carbon.36.1671,PhysRev.134.A453}).
Although some amount of crack is possibly introduced in the mechanical exfoliation process,
these results indicate that the quality of our sample is still reasonable even after the exfoliation process.
In this study, by comparing the critical magnetic field $B_c$ for different thickness samples,
we found that the magnetic-field-induced phase becomes unstable for thin-film samples.

 \section{EXPERIMENTAL METHODS}

Thin-film graphite samples were obtained by mechanical exfoliation from Kish graphite crystals and were transferred onto the silicon substrate, in the same manner as the original graphene preparation~\cite{Science.306.666,Nature.438.197}.
Here, insulating silicon substrates were utilized in order to avoid heating by the eddy current under the pulsed magnetic field.
The thickness of each microcrystal on the substrate was identified by atomic-force microscopy,
in which we selected the flat surface samples.
The typical dimensions of the microcrystal were $50\times 50\times 0.1 \;\mu$m${^3}$.
The electrical contacts for in-plane resistance measurements were formed by standard electron-beam lithography and vacuum evaporation of gold.

High magnetic fields were generated by our portable nondestructive pulse magnet system,
which consists of home-wound coil cooled by liquid nitrogen,
a capacitor bank with a maximum charge energy of 20 kJ,
and a helium cryostat with a lowest temperature of 1.6 K.
The highest magnetic field reached $B \simeq 40$ T at a duration of 10 ms.
Because the time dependence of the magnetic field was very steep and noisy in the ascending branch,
we only show the descending branch (see Appendix~\ref{AppendixNumLockin} in detail.)

The in-plane resistance ($R$) was measured by applying a small ac electric current ($I_{\textrm{ac}}$) at a frequency of 25 kHz under the magnetic fields ($B$) along the $c$-axis (perpendicular to the plate) at low temperatures ($T$).
The resistance was determined by the numerical lock-in technique (see Appendix~\ref{AppendixNumLockin} in detail.)


 \section{RESULTS}
\begin{figure}
\centering
\includegraphics[width=8cm]{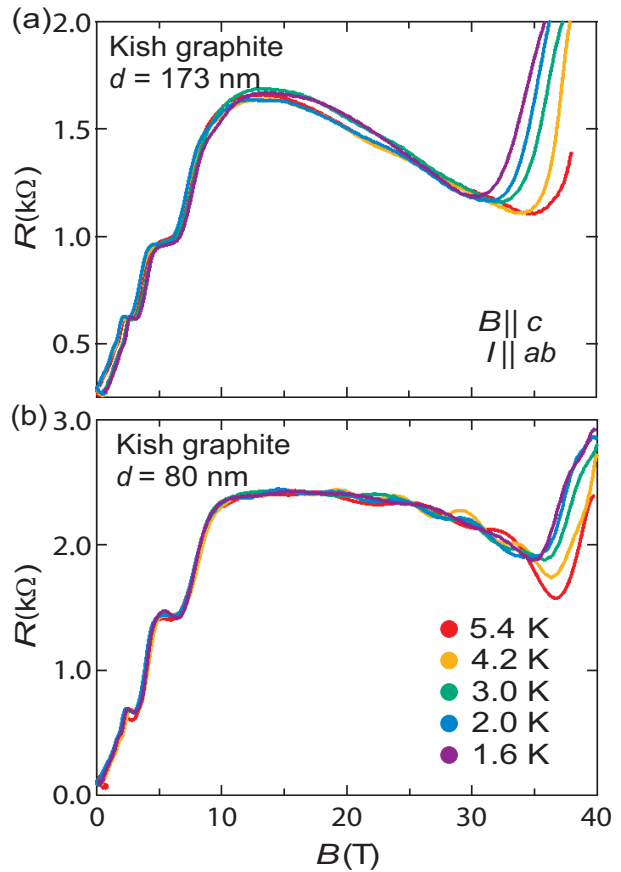}
\caption{In-plane resistance as a function of magnetic field along the $c$ axis in (a) 173-nm-thick and (b) 80-nm-thick graphite at $T = 1.6, 2.0, 3.0, 4.2$, and 5.4 K. Several dip structures up to 10 T are Shubnikov-de Haas oscillations. In both samples, the critical magnetic field of the semimetal-insulator transition increases with elevating temperatures. Critical fields in thinner samples are higher, and show small temperature dependence.}
\label{f2}
\end{figure}

\begin{figure}
\centering
\includegraphics[width=8cm]{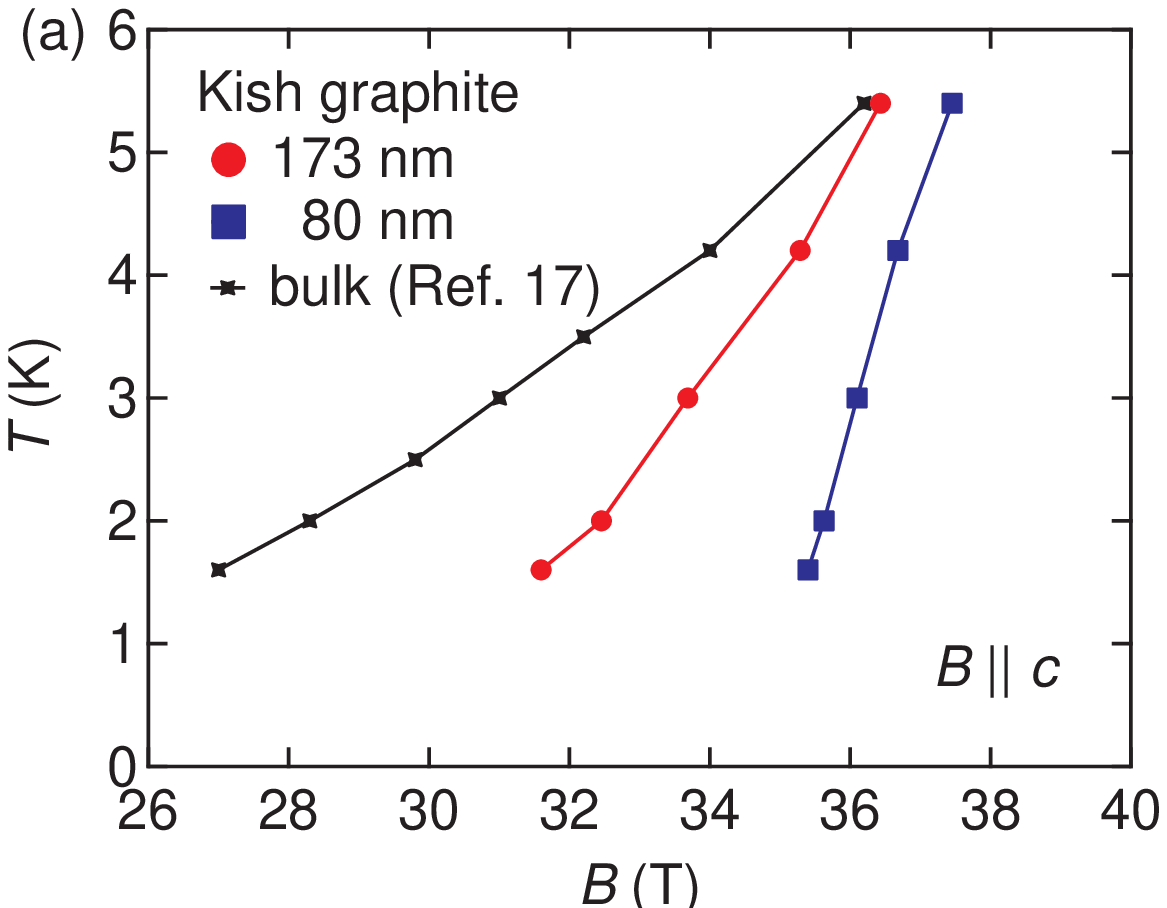}
\includegraphics[width=8cm]{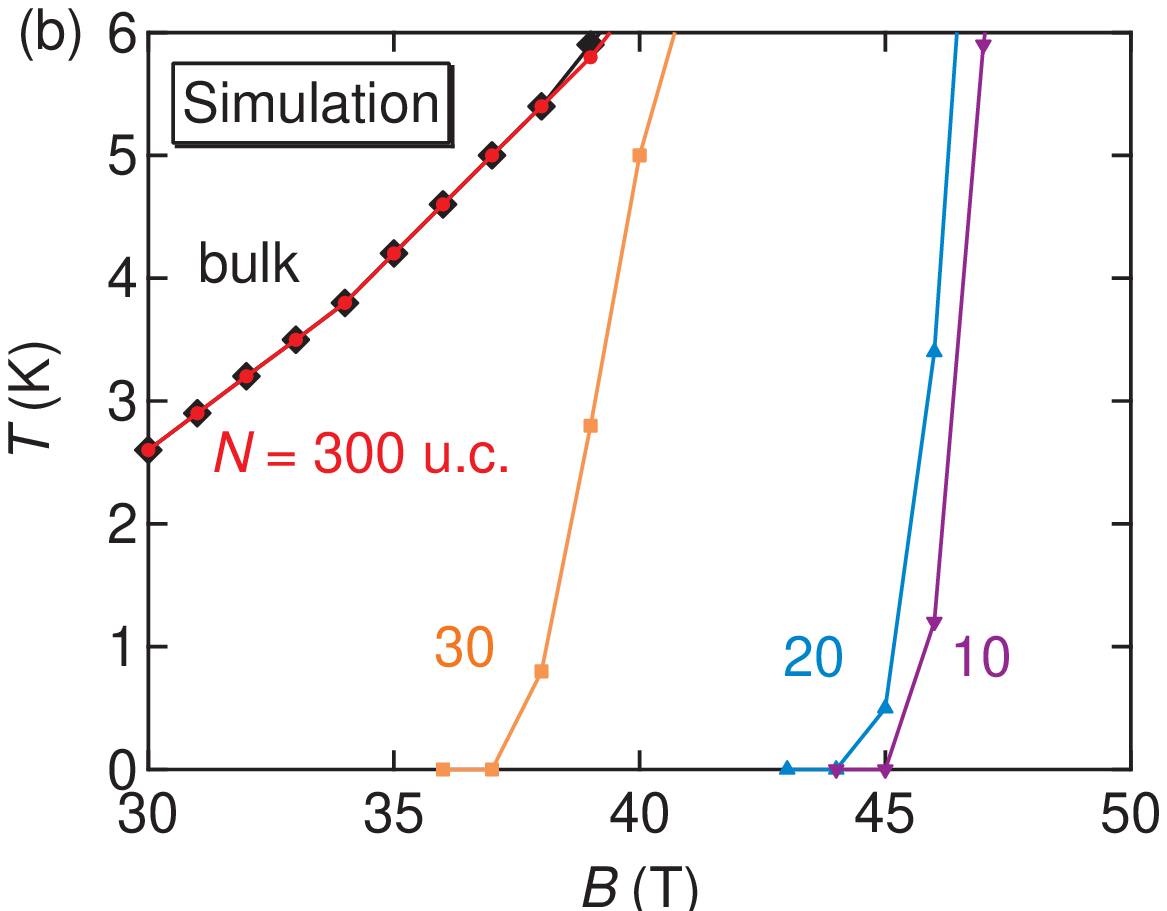}
\caption{(a) Phase boundary of the semimetal-insulator transition in $B-T$ plane for each graphite sample, obtained from Fig.~\ref{f2}. The bulk line is taken from Ref.~[\onlinecite{JPhysSocJpn.68.1300}]. By reducing sample thickness, the phase boundary shifts to higher magnetic fields, and the slope becomes steep. (b) Simulated phase boundaries for several thicknesses. Two characteristics of the thinner system are qualitatively reproduced.}
\label{f3}
\end{figure}

Figures \ref{f2}
(a) and \ref{f2}(b) are the magnetic-field dependences of the in-plane resistance at several temperatures in samples with $d = 173$ nm and 80 nm, respectively.
Both samples show trends similar to those of the bulk sample.
Namely,
a large magnetoresistance appears up to 10 T, concomitant with clear Shubnikov de-Haas oscillations,
followed by a negative magnetoresistance between 10 T and 30 T.
A sharp transition to the insulating state can be observed at around 30 T.
These results indicate that both samples can be viewed as three-dimensional systems.
In fact, mono- and bilayer graphenes show different sequences of the Shubnikov de-Haas oscillations, and the semimetal-insulator transition is absent at around 30 T (Ref.~[\onlinecite{NewJPhys.12.083006}]).
With decreasing temperatures, 
the transition rapidly shifts to lower magnetic fields.
This temperature dependence of the transition is qualitatively the same as that of the bulk result~\cite{PhysicaB.256.621}. 
On the other hand, we can see some differences between the two samples.
In the thinner sample, (i) the value of the critical magnetic field $B_c$ shifts higher, and (ii) the temperature dependence of $B_c$ becomes small.
These characteristics are clearly visualized in the $B-T$ phase diagram,
as shown in Fig.~\ref{f3}(a).
For comparison,
that of the bulk system is also shown,
and is taken from Ref.~[\onlinecite{JPhysSocJpn.68.1300}].
When the thickness is reduced,
the phase boundary line between the semimetal and insulating states (i) shifts to higher magnetic fields,
and (ii) the slope of it becomes steeper.
The second trend is in stark contrast to the phase diagram found in neutron irradiated graphite,
where the boundary almost shifts to higher fields in a parallel manner.
The difference probably comes from an introduction of disorders and charge carriers.
We note that a previous report for a 130-nm sample~\cite{CurrApplPhys.7.338} does not contradict our results,
although the applied magnetic fields are not sufficient to determine the transition in that measurement.
Recently, the transition was observed at $B = 38$ T in the highly-oriented pyrolytic graphite (HOPG) sample with $d = 35$ nm at $T = 4.2$ K.
This result is consistent with our phase diagram.

 \section{DISCUSSION}
The presence of thickness dependence implies that an ordered state along the out-of-plane direction evolves.
To examine whether this ordered state is attributable to the formation of the density-wave state,
we calculated the thickness dependence of $B_c$ in the simple density-wave state model~\cite{RevModPhys.60.1129}.
In the case of a quasi-quantum limit [only four spin-split Landau subbands of ($n=0,\uparrow \downarrow$) and ($n=-1,\uparrow \downarrow$) are at the Fermi level],
the density-response function $\chi (\textbf{q}) = \rho (\textbf{q})/V (\textbf{q})$ can be evaluated by
\begin{eqnarray}
	\lefteqn{ \chi^{(0)}(q_x=0, q_y=0, q_z)}	\nonumber \\
	& = & \frac{1}{2\pi l^2} \sum_{k_z}
	\frac{f(E_{0\uparrow} (k_z+q_z))-f(E_{0\uparrow} (k_z))}{E_{0\uparrow} (k_z) - E_{0\uparrow} (k_z+q_z)}.
	\label{eq:chi0}
\end{eqnarray}
Here,
$\rho (\textbf{q})$ and $V (\textbf{q})$ are the Fourier components of the carrier density and perturbation potential, respectively,
$E_{0\uparrow}$ denotes the ($n = 0,\uparrow$) Landau subband energy dispersion,
$f(E)$ is the Fermi-Dirac distribution function,
$l = \sqrt{\hbar/eB}$ is the magnetic length,
$\hbar=h/2\pi$ is Planck's constant divided by $2\pi$,
and $e$ is the elementary charge.
According to Ref.~[\onlinecite{JPhysSocJpn.50.725}],
the ($n = 0, \uparrow$) Landau subband is relevant for the density-wave transition.
Therefore, 
we focus on the ($n = 0, \uparrow$) Landau subband,
and for simplicity, the Fermi energy is fixed to the value of the bulk system,
regardless of the thickness.
The energy dispersion of the subband $E_{0\uparrow} (k_z)$ is calculated by the Slonczewski-Weiss-McClure model~\cite{PhysRev.109.272,PhysRev.119.606} with $\gamma_3=0$ in Ref.~[\onlinecite{JPhysSocJpn.17.808}],
as this term is not effective in a high magnetic field.
The condition for the density-wave transition is that $\max\left[\chi^{(0)}\right] $ reaches a critical value $1/\tilde{u}$ (Ref.~[\onlinecite{JPhysSocJpn.50.725}]),
where $\tilde{u}$ is an effective exchange interaction.
In a bulk system, i.e., in the limit of the continuous $k_z$,
we can easily evaluate Eq. (\ref{eq:chi0}) by substituting the summation in the integral.
As a result,
we obtain the so-called ``$2k_F$ instability,'' 
namely, $\chi^{(0)}$ divergently increases at $q_z = 2k_F$,
and the peak rapidly decreases with elevating temperatures.
On the other hand, in a thin-film system,
as $k_z$ is discrete with a spacing of $\Delta k_z = 2\pi /d$,
we directly sum all possible $k_z$ and $k_z+q_z$ pairs at each $q_z$ in Eq. (\ref{eq:chi0}).
We note that only an integer multiple of $\Delta k_z$ is allowed for $q_z$,
so $q_z$ is also discrete.
In the $N = 300$ unit cell (u.c.)-thick system ($N=k_z/\Delta k_z=d/c$, where $c$ is the lattice constant along the $c$-axis), 
the results are quite similar to those in the bulk, as it is still thick enough.
However, in thinner cases such as $N = 30, 20$, and 10-u.c. systems,
the values at the peak become smaller.
In addition, the temperature dependence of the peak height becomes progressively smaller by reducing the thickness (see Appendix~\ref{Appendixchi0calc}).
These two features mean that if we assume the critical condition $\max\left[\chi^{(0)}(T,B)\right] = 1/\tilde{u}$ is thickness independent,
the density-wave transition should occur at lower temperatures in thinner systems in some fixed magnetic fields.
By determining some adequate value of $1/\tilde{u}$,
the simulated phase diagram is depicted as Fig.~\ref{f3}(b).
Figure \ref{f3}(b)
qualitatively reproduces the trend of the phase boundary pointed out in the experimental phase diagram of Fig.~\ref{f3}(a) [see (i) and (ii) above].
Taking into account the agreement of the characteristics of the phase boundary,
we strongly suggest that the insulating state that appeared above $B \simeq 30$ T in graphite is the density-wave state.
Note that we do not confirm whether it is the \textit{valley}-density-wave state.
Although our simulation is based on the band dispersion of the Slonczewski-Weiss-McClure model~\cite{PhysRev.109.272,PhysRev.119.606},
the conclusion is not affected by the details of the band structure, as shown in the following discussion.

To look into the quantum size effect on the density-wave state,
we discuss the mechanism for the shift of $B_c$ and the small temperature dependence.
The first feature, (i) the shift of $B_c$, is attributable to the boundary condition for the density-wave state.
If we compare thick and thin samples at some fixed magnetic field around 30 T,
the thick sample has a pair of $k_z$ and $k_z + q_z$ just at the Fermi level,
while in the thin sample, such pairs cannot be found in some cases owing to the sparse $k_z$ [see Fig.~\ref{f1}(b)].
This means that the thin-film system needs to tune the magnetic field to find a pair of $k_z$ and $k_z + q_z$.
Because the value of $\max\left[\chi^{(0)}(T,B)\right]$ monotonically increases with $B$, as indicated by Eq.~(\ref{eq:chi0}),
the thin-film system needs a higher magnetic field to find a pair to achieve $\max\left[\chi^{(0)}(T,B)\right] \geq 1/\tilde{u}$.
In real space,
this feature corresponds to the formation of the density wave with a fixed-end boundary condition.
In the thick system,
where the boundary condition is not relevant,
the formation of the density wave is not restricted by the characteristic length of the density wave $\sim \pi/k_F$.
On the other hand,
as the node position of the density wave is expected to come at the boundary,
a mismatch of $d$ and $\sim \pi/k_F$ will make it difficult for the system to enter the density-wave state.
In fact,
the simulated $B_c (T)$ nonmonotonically behaves in the fixed $\varepsilon_F$ calculation [not shown in Fig.~\ref{f3}(b)],
although in reality the Fermi energy will go up and down as the magnetic fields increase.
The second feature, (II) the small temperature dependence of $B_c(T)$, originates from the sparseness of the states along the energy direction in the Landau subbands, instead of that along the $k_z$ direction.
Because the energy spectra are no longer continuous by the quantum size effect,
the distribution function becomes irrelevant to the system.
More specifically,
if we draw horizontal bars with two different widths of $k_B T$, as indicated in Fig.~\ref{f1}(b) by light blue (lower temperature) and light red (higher temperature),
the number of points on $E(k_z)$ overlaid by these bars are almost the same in the thin-film system,
while remarkably different in the thick-enough system
owing to the different energy spacing.
Hence, the condition satisfying the density-wave transition is not affected by the temperature,
resulting in the steep phase boundary in Fig.~\ref{f3}(b).

Finally, the threshold of the thickness is discussed, below which the quantum size effect becomes relevant.
Surprisingly,
the relatively thick system of $d=173$ nm already deviates from the bulk phase boundary in our experimental results,
but this value is on the same order of 70 nm, the rough estimation mentioned above.
Our density-response function calculation also supports this result.
Hence, we can safely attribute the phase boundary shift to the quantum size effect,
although a factor of difference remains.
In fact, our density-response function calculation shows that
a 300-u.c. system, which corresponds to $d\simeq 200$ nm, shows the same result as the bulk one.
This quantitative refinement is required by a modification of the nesting vector or a selection of subband.
Further investigation is expected for the quantitative agreement.

 \section{CONCLUSION}

In conclusion,
we observed a thickness-dependent electronic phase transition at $B\simeq 30$ T in graphite.
The transition in thin-film graphite on silicon substrate was detected by the in-plane transport under a pulsed magnetic field.
The thickness dependence of the transition indicates that the ordered state along the out-of-plane direction evolves.
In contrast to bulk graphite,
the critical magnetic field in thin-film graphite shifts higher with reduced temperature dependence.
These features are understood by the quantum size effect on the density-wave transition,
and the phase diagram is in reasonably good agreement with the simulated one based on the density-wave state.
As a result, we strongly suggest that the insulating state appearing at $B\simeq 30$ T is the density-wave state.
This thinning approach, which controls the phase transition through level spacing without introducing defects or carriers, will help us understand the entire phase diagram of graphite.

\begin{acknowledgments}
The authors are grateful to Professor Y. Takada, Professor M. Tokunaga, Professor Y. Iye, Professor H. Yaguchi, and Professor R. Shindo for valuable discussions.
This work was partially supported by JSPS KAKENHI Grant No. JP25107003.
\end{acknowledgments}

\bibliography{bibliography}

\begin{thebibliography}{32}%
\makeatletter
\providecommand \@ifxundefined [1]{%
 \@ifx{#1\undefined}
}%
\providecommand \@ifnum [1]{%
 \ifnum #1\expandafter \@firstoftwo
 \else \expandafter \@secondoftwo
 \fi
}%
\providecommand \@ifx [1]{%
 \ifx #1\expandafter \@firstoftwo
 \else \expandafter \@secondoftwo
 \fi
}%
\providecommand \natexlab [1]{#1}%
\providecommand \enquote  [1]{``#1''}%
\providecommand \bibnamefont  [1]{#1}%
\providecommand \bibfnamefont [1]{#1}%
\providecommand \citenamefont [1]{#1}%
\providecommand \href@noop [0]{\@secondoftwo}%
\providecommand \href [0]{\begingroup \@sanitize@url \@href}%
\providecommand \@href[1]{\@@startlink{#1}\@@href}%
\providecommand \@@href[1]{\endgroup#1\@@endlink}%
\providecommand \@sanitize@url [0]{\catcode `\\12\catcode `\$12\catcode
  `\&12\catcode `\#12\catcode `\^12\catcode `\_12\catcode `\%12\relax}%
\providecommand \@@startlink[1]{}%
\providecommand \@@endlink[0]{}%
\providecommand \url  [0]{\begingroup\@sanitize@url \@url }%
\providecommand \@url [1]{\endgroup\@href {#1}{\urlprefix }}%
\providecommand \urlprefix  [0]{URL }%
\providecommand \Eprint [0]{\href }%
\providecommand \doibase [0]{http://dx.doi.org/}%
\providecommand \selectlanguage [0]{\@gobble}%
\providecommand \bibinfo  [0]{\@secondoftwo}%
\providecommand \bibfield  [0]{\@secondoftwo}%
\providecommand \translation [1]{[#1]}%
\providecommand \BibitemOpen [0]{}%
\providecommand \bibitemStop [0]{}%
\providecommand \bibitemNoStop [0]{.\EOS\space}%
\providecommand \EOS [0]{\spacefactor3000\relax}%
\providecommand \BibitemShut  [1]{\csname bibitem#1\endcsname}%
\let\auto@bib@innerbib\@empty
\bibitem [{\citenamefont {Tanuma}\ \emph {et~al.}(1981)\citenamefont {Tanuma},
  \citenamefont {Inada}, \citenamefont {Furukawa}, \citenamefont {Takahashi},
  \citenamefont {Iye},\ and\ \citenamefont {Onuki}}]{Tanuma1981}%
  \BibitemOpen
  \bibfield  {author} {\bibinfo {author} {\bibfnamefont {S.}~\bibnamefont
  {Tanuma}}, \bibinfo {author} {\bibfnamefont {R.}~\bibnamefont {Inada}},
  \bibinfo {author} {\bibfnamefont {A.}~\bibnamefont {Furukawa}}, \bibinfo
  {author} {\bibfnamefont {O.}~\bibnamefont {Takahashi}}, \bibinfo {author}
  {\bibfnamefont {Y.}~\bibnamefont {Iye}}, \ and\ \bibinfo {author}
  {\bibfnamefont {Y.}~\bibnamefont {Onuki}},\ }\href@noop {} {\emph {\bibinfo
  {title} {Physics in High Magnetic Fields}}},\ edited by\ \bibinfo {editor}
  {\bibfnamefont {S.}~\bibnamefont {Chikazumi}}\ and\ \bibinfo {editor}
  {\bibfnamefont {N.}~\bibnamefont {Miura}}\ (\bibinfo  {publisher}
  {Springer},\ \bibinfo {address} {Berlin},\ \bibinfo {year} {1981})\ p.\
  \bibinfo {pages} {316}\BibitemShut {NoStop}%
\bibitem [{\citenamefont {Fauqu\'e}\ \emph {et~al.}(2013)\citenamefont
  {Fauqu\'e}, \citenamefont {LeBoeuf}, \citenamefont {Vignolle}, \citenamefont
  {Nardone}, \citenamefont {Proust},\ and\ \citenamefont
  {Behnia}}]{PhysRevLett.110.266601}%
  \BibitemOpen
  \bibfield  {author} {\bibinfo {author} {\bibfnamefont {B.}~\bibnamefont
  {Fauqu\'e}}, \bibinfo {author} {\bibfnamefont {D.}~\bibnamefont {LeBoeuf}},
  \bibinfo {author} {\bibfnamefont {B.}~\bibnamefont {Vignolle}}, \bibinfo
  {author} {\bibfnamefont {M.}~\bibnamefont {Nardone}}, \bibinfo {author}
  {\bibfnamefont {C.}~\bibnamefont {Proust}}, \ and\ \bibinfo {author}
  {\bibfnamefont {K.}~\bibnamefont {Behnia}},\ }\href@noop {} {\bibfield
  {journal} {\bibinfo  {journal} {Phys. Rev. Lett.}\ }\textbf {\bibinfo
  {volume} {110}},\ \bibinfo {pages} {266601} (\bibinfo {year}
  {2013})}\BibitemShut {NoStop}%
\bibitem [{\citenamefont {Akiba}\ \emph {et~al.}(2015)\citenamefont {Akiba},
  \citenamefont {Miyake}, \citenamefont {Yaguchi}, \citenamefont {Matsuo},
  \citenamefont {Kindo},\ and\ \citenamefont
  {Tokunaga}}]{JPhysSocJpn.84.054709}%
  \BibitemOpen
  \bibfield  {author} {\bibinfo {author} {\bibfnamefont {K.}~\bibnamefont
  {Akiba}}, \bibinfo {author} {\bibfnamefont {A.}~\bibnamefont {Miyake}},
  \bibinfo {author} {\bibfnamefont {H.}~\bibnamefont {Yaguchi}}, \bibinfo
  {author} {\bibfnamefont {A.}~\bibnamefont {Matsuo}}, \bibinfo {author}
  {\bibfnamefont {K.}~\bibnamefont {Kindo}}, \ and\ \bibinfo {author}
  {\bibfnamefont {M.}~\bibnamefont {Tokunaga}},\ }\href@noop {} {\bibfield
  {journal} {\bibinfo  {journal} {J. Phys. Soc. Jpn.}\ }\textbf {\bibinfo
  {volume} {84}},\ \bibinfo {pages} {054709} (\bibinfo {year}
  {2015})}\BibitemShut {NoStop}%
\bibitem [{\citenamefont {Kopelevich}\ \emph {et~al.}(2009)\citenamefont
  {Kopelevich}, \citenamefont {Raquet}, \citenamefont {Goiran}, \citenamefont
  {Escoffier}, \citenamefont {da~Silva}, \citenamefont {Medina~Pantoja},
  \citenamefont {Luk'yanchuk}, \citenamefont {Sinchenko},\ and\ \citenamefont
  {Monceau}}]{PhysRevLett.103.116802}%
  \BibitemOpen
  \bibfield  {author} {\bibinfo {author} {\bibfnamefont {Y.}~\bibnamefont
  {Kopelevich}}, \bibinfo {author} {\bibfnamefont {B.}~\bibnamefont {Raquet}},
  \bibinfo {author} {\bibfnamefont {M.}~\bibnamefont {Goiran}}, \bibinfo
  {author} {\bibfnamefont {W.}~\bibnamefont {Escoffier}}, \bibinfo {author}
  {\bibfnamefont {R.~R.}\ \bibnamefont {da~Silva}}, \bibinfo {author}
  {\bibfnamefont {J.~C.}\ \bibnamefont {Medina~Pantoja}}, \bibinfo {author}
  {\bibfnamefont {I.~A.}\ \bibnamefont {Luk'yanchuk}}, \bibinfo {author}
  {\bibfnamefont {A.}~\bibnamefont {Sinchenko}}, \ and\ \bibinfo {author}
  {\bibfnamefont {P.}~\bibnamefont {Monceau}},\ }\href@noop {} {\bibfield
  {journal} {\bibinfo  {journal} {Phys. Rev. Lett.}\ }\textbf {\bibinfo
  {volume} {103}},\ \bibinfo {pages} {116802} (\bibinfo {year}
  {2009})}\BibitemShut {NoStop}%
\bibitem [{\citenamefont {Arnold}\ \emph {et~al.}(2017)\citenamefont {Arnold},
  \citenamefont {Isidori}, \citenamefont {Kampert}, \citenamefont {Yager},
  \citenamefont {Eschrig},\ and\ \citenamefont
  {Saunders}}]{PhysRevLett.119.136601}%
  \BibitemOpen
  \bibfield  {author} {\bibinfo {author} {\bibfnamefont {F.}~\bibnamefont
  {Arnold}}, \bibinfo {author} {\bibfnamefont {A.}~\bibnamefont {Isidori}},
  \bibinfo {author} {\bibfnamefont {E.}~\bibnamefont {Kampert}}, \bibinfo
  {author} {\bibfnamefont {B.}~\bibnamefont {Yager}}, \bibinfo {author}
  {\bibfnamefont {M.}~\bibnamefont {Eschrig}}, \ and\ \bibinfo {author}
  {\bibfnamefont {J.}~\bibnamefont {Saunders}},\ }\href@noop {} {\bibfield
  {journal} {\bibinfo  {journal} {Phys. Rev. Lett.}\ }\textbf {\bibinfo
  {volume} {119}},\ \bibinfo {pages} {136601} (\bibinfo {year}
  {2017})}\BibitemShut {NoStop}%
\bibitem [{\citenamefont {Zhu}\ \emph {et~al.}(2017)\citenamefont {Zhu},
  \citenamefont {McDonald}, \citenamefont {Shekhter}, \citenamefont {Ramshaw},
  \citenamefont {Modic}, \citenamefont {Balakirev},\ and\ \citenamefont
  {Harrison}}]{SciRep.7.1733}%
  \BibitemOpen
  \bibfield  {author} {\bibinfo {author} {\bibfnamefont {Z.}~\bibnamefont
  {Zhu}}, \bibinfo {author} {\bibfnamefont {R.~D.}\ \bibnamefont {McDonald}},
  \bibinfo {author} {\bibfnamefont {A.}~\bibnamefont {Shekhter}}, \bibinfo
  {author} {\bibfnamefont {B.~J.}\ \bibnamefont {Ramshaw}}, \bibinfo {author}
  {\bibfnamefont {K.~A.}\ \bibnamefont {Modic}}, \bibinfo {author}
  {\bibfnamefont {F.~F.}\ \bibnamefont {Balakirev}}, \ and\ \bibinfo {author}
  {\bibfnamefont {N.}~\bibnamefont {Harrison}},\ }\href@noop {} {\bibfield
  {journal} {\bibinfo  {journal} {Sci. Rep.}\ }\textbf {\bibinfo {volume}
  {7}},\ \bibinfo {pages} {1733} (\bibinfo {year} {2017})}\BibitemShut
  {NoStop}%
\bibitem [{\citenamefont {Camargo}\ and\ \citenamefont
  {Escoffier}(2018)}]{CamargoEscoffier2018}%
  \BibitemOpen
  \bibfield  {author} {\bibinfo {author} {\bibfnamefont {B.~C.}\ \bibnamefont
  {Camargo}}\ and\ \bibinfo {author} {\bibfnamefont {W.}~\bibnamefont
  {Escoffier}},\ }\href@noop {} {\enquote {\bibinfo {title} {Taming the
  magnetoresistance anomaly in graphite},}\ } (\bibinfo {year} {2018}),\
  \bibinfo {note} {preprint}\BibitemShut {NoStop}%
\bibitem [{\citenamefont {Slonczewski}\ and\ \citenamefont
  {Weiss}(1958)}]{PhysRev.109.272}%
  \BibitemOpen
  \bibfield  {author} {\bibinfo {author} {\bibfnamefont {J.~C.}\ \bibnamefont
  {Slonczewski}}\ and\ \bibinfo {author} {\bibfnamefont {P.~R.}\ \bibnamefont
  {Weiss}},\ }\href@noop {} {\bibfield  {journal} {\bibinfo  {journal} {Phys.
  Rev.}\ }\textbf {\bibinfo {volume} {109}},\ \bibinfo {pages} {272} (\bibinfo
  {year} {1958})}\BibitemShut {NoStop}%
\bibitem [{\citenamefont {McClure}(1960)}]{PhysRev.119.606}%
  \BibitemOpen
  \bibfield  {author} {\bibinfo {author} {\bibfnamefont {J.~W.}\ \bibnamefont
  {McClure}},\ }\href@noop {} {\bibfield  {journal} {\bibinfo  {journal} {Phys.
  Rev.}\ }\textbf {\bibinfo {volume} {119}},\ \bibinfo {pages} {606} (\bibinfo
  {year} {1960})}\BibitemShut {NoStop}%
\bibitem [{\citenamefont {Inoue}(1962)}]{JPhysSocJpn.17.808}%
  \BibitemOpen
  \bibfield  {author} {\bibinfo {author} {\bibfnamefont {M.}~\bibnamefont
  {Inoue}},\ }\href@noop {} {\bibfield  {journal} {\bibinfo  {journal} {J.
  Phys. Soc. Jpn.}\ }\textbf {\bibinfo {volume} {17}},\ \bibinfo {pages} {808}
  (\bibinfo {year} {1962})}\BibitemShut {NoStop}%
\bibitem [{\citenamefont {Yoshioka}\ and\ \citenamefont
  {Fukuyama}(1981)}]{JPhysSocJpn.50.725}%
  \BibitemOpen
  \bibfield  {author} {\bibinfo {author} {\bibfnamefont {D.}~\bibnamefont
  {Yoshioka}}\ and\ \bibinfo {author} {\bibfnamefont {H.}~\bibnamefont
  {Fukuyama}},\ }\href@noop {} {\bibfield  {journal} {\bibinfo  {journal} {J.
  Phys. Soc. Jpn.}\ }\textbf {\bibinfo {volume} {50}},\ \bibinfo {pages} {725}
  (\bibinfo {year} {1981})}\BibitemShut {NoStop}%
\bibitem [{\citenamefont {Takahashi}\ and\ \citenamefont
  {Takada}(1994)}]{PhysicaB.201.384}%
  \BibitemOpen
  \bibfield  {author} {\bibinfo {author} {\bibfnamefont {K.}~\bibnamefont
  {Takahashi}}\ and\ \bibinfo {author} {\bibfnamefont {Y.}~\bibnamefont
  {Takada}},\ }\href@noop {} {\bibfield  {journal} {\bibinfo  {journal}
  {Physica B: Condensed Matter}\ }\textbf {\bibinfo {volume} {201}},\ \bibinfo
  {pages} {384 } (\bibinfo {year} {1994})}\BibitemShut {NoStop}%
\bibitem [{\citenamefont {Sugihara}(1984)}]{PhysRevB.29.6722}%
  \BibitemOpen
  \bibfield  {author} {\bibinfo {author} {\bibfnamefont {K.}~\bibnamefont
  {Sugihara}},\ }\href@noop {} {\bibfield  {journal} {\bibinfo  {journal}
  {Phys. Rev. B}\ }\textbf {\bibinfo {volume} {29}},\ \bibinfo {pages} {6722}
  (\bibinfo {year} {1984})}\BibitemShut {NoStop}%
\bibitem [{\citenamefont {Takada}\ and\ \citenamefont
  {Goto}(1998)}]{JPhysCondensMatter.10.11315}%
  \BibitemOpen
  \bibfield  {author} {\bibinfo {author} {\bibfnamefont {Y.}~\bibnamefont
  {Takada}}\ and\ \bibinfo {author} {\bibfnamefont {H.}~\bibnamefont {Goto}},\
  }\href@noop {} {\bibfield  {journal} {\bibinfo  {journal} {J. Phys.: Condens.
  Matter}\ }\textbf {\bibinfo {volume} {10}},\ \bibinfo {pages} {11315}
  (\bibinfo {year} {1998})}\BibitemShut {NoStop}%
\bibitem [{\citenamefont {Iye}\ and\ \citenamefont
  {Dresselhaus}(1985)}]{PhysRevLett.54.1182}%
  \BibitemOpen
  \bibfield  {author} {\bibinfo {author} {\bibfnamefont {Y.}~\bibnamefont
  {Iye}}\ and\ \bibinfo {author} {\bibfnamefont {G.}~\bibnamefont
  {Dresselhaus}},\ }\href@noop {} {\bibfield  {journal} {\bibinfo  {journal}
  {Phys. Rev. Lett.}\ }\textbf {\bibinfo {volume} {54}},\ \bibinfo {pages}
  {1182} (\bibinfo {year} {1985})}\BibitemShut {NoStop}%
\bibitem [{\citenamefont {Yaguchi}\ \emph
  {et~al.}(1999{\natexlab{a}})\citenamefont {Yaguchi}, \citenamefont
  {Takamasu}, \citenamefont {Iye},\ and\ \citenamefont
  {Miura}}]{JPhysSocJpn.68.181}%
  \BibitemOpen
  \bibfield  {author} {\bibinfo {author} {\bibfnamefont {H.}~\bibnamefont
  {Yaguchi}}, \bibinfo {author} {\bibfnamefont {T.}~\bibnamefont {Takamasu}},
  \bibinfo {author} {\bibfnamefont {Y.}~\bibnamefont {Iye}}, \ and\ \bibinfo
  {author} {\bibfnamefont {N.}~\bibnamefont {Miura}},\ }\href@noop {}
  {\bibfield  {journal} {\bibinfo  {journal} {J. Phys. Soc. Jpn.}\ }\textbf
  {\bibinfo {volume} {68}},\ \bibinfo {pages} {181} (\bibinfo {year}
  {1999}{\natexlab{a}})}\BibitemShut {NoStop}%
\bibitem [{\citenamefont {Yaguchi}\ \emph
  {et~al.}(1999{\natexlab{b}})\citenamefont {Yaguchi}, \citenamefont {Iye},
  \citenamefont {Takamasu}, \citenamefont {Miura},\ and\ \citenamefont
  {Iwata}}]{JPhysSocJpn.68.1300}%
  \BibitemOpen
  \bibfield  {author} {\bibinfo {author} {\bibfnamefont {H.}~\bibnamefont
  {Yaguchi}}, \bibinfo {author} {\bibfnamefont {Y.}~\bibnamefont {Iye}},
  \bibinfo {author} {\bibfnamefont {T.}~\bibnamefont {Takamasu}}, \bibinfo
  {author} {\bibfnamefont {N.}~\bibnamefont {Miura}}, \ and\ \bibinfo {author}
  {\bibfnamefont {T.}~\bibnamefont {Iwata}},\ }\href@noop {} {\bibfield
  {journal} {\bibinfo  {journal} {J. Phys. Soc. Jpn.}\ }\textbf {\bibinfo
  {volume} {68}},\ \bibinfo {pages} {1300} (\bibinfo {year}
  {1999}{\natexlab{b}})}\BibitemShut {NoStop}%
\bibitem [{\citenamefont {Yaguchi}\ \emph {et~al.}(2001)\citenamefont
  {Yaguchi}, \citenamefont {Singleton},\ and\ \citenamefont
  {Iwata}}]{PhysicaB.298.546}%
  \BibitemOpen
  \bibfield  {author} {\bibinfo {author} {\bibfnamefont {H.}~\bibnamefont
  {Yaguchi}}, \bibinfo {author} {\bibfnamefont {J.}~\bibnamefont {Singleton}},
  \ and\ \bibinfo {author} {\bibfnamefont {T.}~\bibnamefont {Iwata}},\
  }\href@noop {} {\bibfield  {journal} {\bibinfo  {journal} {Physica B:
  Condensed Matter}\ }\textbf {\bibinfo {volume} {298}},\ \bibinfo {pages} {546
  } (\bibinfo {year} {2001})}\BibitemShut {NoStop}%
\bibitem [{\citenamefont {Yaguchi}\ and\ \citenamefont
  {Singleton}(2009{\natexlab{a}})}]{JPhysConfSer.150.022099}%
  \BibitemOpen
  \bibfield  {author} {\bibinfo {author} {\bibfnamefont {H.}~\bibnamefont
  {Yaguchi}}\ and\ \bibinfo {author} {\bibfnamefont {J.}~\bibnamefont
  {Singleton}},\ }\href@noop {} {\bibfield  {journal} {\bibinfo  {journal} {J.
  Phys.: Conf. Ser.}\ }\textbf {\bibinfo {volume} {150}},\ \bibinfo {pages}
  {022099} (\bibinfo {year} {2009}{\natexlab{a}})}\BibitemShut {NoStop}%
\bibitem [{\citenamefont {Yaguchi}\ and\ \citenamefont
  {Singleton}(2009{\natexlab{b}})}]{JPhysCondensMattter.21.344207}%
  \BibitemOpen
  \bibfield  {author} {\bibinfo {author} {\bibfnamefont {H.}~\bibnamefont
  {Yaguchi}}\ and\ \bibinfo {author} {\bibfnamefont {J.}~\bibnamefont
  {Singleton}},\ }\href@noop {} {\bibfield  {journal} {\bibinfo  {journal} {J.
  Phys.: Condens. Matter}\ }\textbf {\bibinfo {volume} {21}},\ \bibinfo {pages}
  {344207} (\bibinfo {year} {2009}{\natexlab{b}})}\BibitemShut {NoStop}%
\bibitem [{\citenamefont {Iye}\ \emph {et~al.}(1984)\citenamefont {Iye},
  \citenamefont {Berglund},\ and\ \citenamefont
  {McNeil}}]{SolidStateCommun.52.975}%
  \BibitemOpen
  \bibfield  {author} {\bibinfo {author} {\bibfnamefont {Y.}~\bibnamefont
  {Iye}}, \bibinfo {author} {\bibfnamefont {P.}~\bibnamefont {Berglund}}, \
  and\ \bibinfo {author} {\bibfnamefont {L.}~\bibnamefont {McNeil}},\
  }\href@noop {} {\bibfield  {journal} {\bibinfo  {journal} {Solid State
  Commun.}\ }\textbf {\bibinfo {volume} {52}},\ \bibinfo {pages} {975 }
  (\bibinfo {year} {1984})}\BibitemShut {NoStop}%
\bibitem [{\citenamefont {Schneider}\ \emph {et~al.}(2009)\citenamefont
  {Schneider}, \citenamefont {Orlita}, \citenamefont {Potemski},\ and\
  \citenamefont {Maude}}]{PhysRevLett.102.166403}%
  \BibitemOpen
  \bibfield  {author} {\bibinfo {author} {\bibfnamefont {J.~M.}\ \bibnamefont
  {Schneider}}, \bibinfo {author} {\bibfnamefont {M.}~\bibnamefont {Orlita}},
  \bibinfo {author} {\bibfnamefont {M.}~\bibnamefont {Potemski}}, \ and\
  \bibinfo {author} {\bibfnamefont {D.~K.}\ \bibnamefont {Maude}},\ }\href@noop
  {} {\bibfield  {journal} {\bibinfo  {journal} {Phys. Rev. Lett.}\ }\textbf
  {\bibinfo {volume} {102}},\ \bibinfo {pages} {166403} (\bibinfo {year}
  {2009})}\BibitemShut {NoStop}%
\bibitem [{\citenamefont {Yaguchi}\ and\ \citenamefont
  {Singleton}(1998)}]{PhysicaB.256.621}%
  \BibitemOpen
  \bibfield  {author} {\bibinfo {author} {\bibfnamefont {H.}~\bibnamefont
  {Yaguchi}}\ and\ \bibinfo {author} {\bibfnamefont {J.}~\bibnamefont
  {Singleton}},\ }\href@noop {} {\bibfield  {journal} {\bibinfo  {journal}
  {Physica B: Condensed Matter}\ }\textbf {\bibinfo {volume} {256}},\ \bibinfo
  {pages} {621 } (\bibinfo {year} {1998})}\BibitemShut {NoStop}%
\bibitem [{\citenamefont {Kaburagi}\ and\ \citenamefont
  {Hishiyama}(1998)}]{Carbon.36.1671}%
  \BibitemOpen
  \bibfield  {author} {\bibinfo {author} {\bibfnamefont {Y.}~\bibnamefont
  {Kaburagi}}\ and\ \bibinfo {author} {\bibfnamefont {Y.}~\bibnamefont
  {Hishiyama}},\ }\href {\doibase
  https://doi.org/10.1016/S0008-6223(98)00163-8} {\bibfield  {journal}
  {\bibinfo  {journal} {Carbon}\ }\textbf {\bibinfo {volume} {36}},\ \bibinfo
  {pages} {1671 } (\bibinfo {year} {1998})}\BibitemShut {NoStop}%
\bibitem [{\citenamefont {Soule}\ \emph {et~al.}(1964)\citenamefont {Soule},
  \citenamefont {McClure},\ and\ \citenamefont {Smith}}]{PhysRev.134.A453}%
  \BibitemOpen
  \bibfield  {author} {\bibinfo {author} {\bibfnamefont {D.~E.}\ \bibnamefont
  {Soule}}, \bibinfo {author} {\bibfnamefont {J.~W.}\ \bibnamefont {McClure}},
  \ and\ \bibinfo {author} {\bibfnamefont {L.~B.}\ \bibnamefont {Smith}},\
  }\href {\doibase 10.1103/PhysRev.134.A453} {\bibfield  {journal} {\bibinfo
  {journal} {Phys. Rev.}\ }\textbf {\bibinfo {volume} {134}},\ \bibinfo {pages}
  {A453} (\bibinfo {year} {1964})}\BibitemShut {NoStop}%
\bibitem [{\citenamefont {Novoselov}\ \emph {et~al.}(2004)\citenamefont
  {Novoselov}, \citenamefont {Geim}, \citenamefont {Morozov}, \citenamefont
  {Jiang}, \citenamefont {Zhang}, \citenamefont {Dubonos}, \citenamefont
  {Grigorieva},\ and\ \citenamefont {Firsov}}]{Science.306.666}%
  \BibitemOpen
  \bibfield  {author} {\bibinfo {author} {\bibfnamefont {K.~S.}\ \bibnamefont
  {Novoselov}}, \bibinfo {author} {\bibfnamefont {A.~K.}\ \bibnamefont {Geim}},
  \bibinfo {author} {\bibfnamefont {S.~V.}\ \bibnamefont {Morozov}}, \bibinfo
  {author} {\bibfnamefont {D.}~\bibnamefont {Jiang}}, \bibinfo {author}
  {\bibfnamefont {Y.}~\bibnamefont {Zhang}}, \bibinfo {author} {\bibfnamefont
  {S.~V.}\ \bibnamefont {Dubonos}}, \bibinfo {author} {\bibfnamefont {I.~V.}\
  \bibnamefont {Grigorieva}}, \ and\ \bibinfo {author} {\bibfnamefont {A.~A.}\
  \bibnamefont {Firsov}},\ }\href@noop {} {\bibfield  {journal} {\bibinfo
  {journal} {Science}\ }\textbf {\bibinfo {volume} {306}},\ \bibinfo {pages}
  {666} (\bibinfo {year} {2004})}\BibitemShut {NoStop}%
\bibitem [{\citenamefont {Novoselov}\ \emph {et~al.}(2005)\citenamefont
  {Novoselov}, \citenamefont {Geim}, \citenamefont {Morozov}, \citenamefont
  {Jiang}, \citenamefont {Katsnelson}, \citenamefont {Grigorieva},
  \citenamefont {Dubonos},\ and\ \citenamefont {Firsov}}]{Nature.438.197}%
  \BibitemOpen
  \bibfield  {author} {\bibinfo {author} {\bibfnamefont {K.~S.}\ \bibnamefont
  {Novoselov}}, \bibinfo {author} {\bibfnamefont {A.~K.}\ \bibnamefont {Geim}},
  \bibinfo {author} {\bibfnamefont {S.~V.}\ \bibnamefont {Morozov}}, \bibinfo
  {author} {\bibfnamefont {D.}~\bibnamefont {Jiang}}, \bibinfo {author}
  {\bibfnamefont {M.~I.}\ \bibnamefont {Katsnelson}}, \bibinfo {author}
  {\bibfnamefont {I.~V.}\ \bibnamefont {Grigorieva}}, \bibinfo {author}
  {\bibfnamefont {S.~V.}\ \bibnamefont {Dubonos}}, \ and\ \bibinfo {author}
  {\bibfnamefont {A.~A.}\ \bibnamefont {Firsov}},\ }\href@noop {} {\bibfield
  {journal} {\bibinfo  {journal} {Nature (London)}\ }\textbf {\bibinfo {volume}
  {438}},\ \bibinfo {pages} {197} (\bibinfo {year} {2005})}\BibitemShut
  {NoStop}%
\bibitem [{\citenamefont {Poumirol}\ \emph {et~al.}(2010)\citenamefont
  {Poumirol}, \citenamefont {Escoffier}, \citenamefont {Kumar}, \citenamefont
  {Goiran}, \citenamefont {Raquet},\ and\ \citenamefont
  {Broto}}]{NewJPhys.12.083006}%
  \BibitemOpen
  \bibfield  {author} {\bibinfo {author} {\bibfnamefont {J.~M.}\ \bibnamefont
  {Poumirol}}, \bibinfo {author} {\bibfnamefont {W.}~\bibnamefont {Escoffier}},
  \bibinfo {author} {\bibfnamefont {A.}~\bibnamefont {Kumar}}, \bibinfo
  {author} {\bibfnamefont {M.}~\bibnamefont {Goiran}}, \bibinfo {author}
  {\bibfnamefont {B.}~\bibnamefont {Raquet}}, \ and\ \bibinfo {author}
  {\bibfnamefont {J.}~\bibnamefont {Broto}},\ }\href@noop {} {\bibfield
  {journal} {\bibinfo  {journal} {New J. Phys.}\ }\textbf {\bibinfo {volume}
  {12}},\ \bibinfo {pages} {083006} (\bibinfo {year} {2010})}\BibitemShut
  {NoStop}%
\bibitem [{\citenamefont {Jobiliong}\ \emph {et~al.}(2007)\citenamefont
  {Jobiliong}, \citenamefont {Park}, \citenamefont {Brooks},\ and\
  \citenamefont {Vasic}}]{CurrApplPhys.7.338}%
  \BibitemOpen
  \bibfield  {author} {\bibinfo {author} {\bibfnamefont {E.}~\bibnamefont
  {Jobiliong}}, \bibinfo {author} {\bibfnamefont {J.}~\bibnamefont {Park}},
  \bibinfo {author} {\bibfnamefont {J.}~\bibnamefont {Brooks}}, \ and\ \bibinfo
  {author} {\bibfnamefont {R.}~\bibnamefont {Vasic}},\ }\href@noop {}
  {\bibfield  {journal} {\bibinfo  {journal} {Curr. Appl. Phys.}\ }\textbf
  {\bibinfo {volume} {7}},\ \bibinfo {pages} {338 } (\bibinfo {year}
  {2007})}\BibitemShut {NoStop}%
\bibitem [{\citenamefont {Gr\"uner}(1988)}]{RevModPhys.60.1129}%
  \BibitemOpen
  \bibfield  {author} {\bibinfo {author} {\bibfnamefont {G.}~\bibnamefont
  {Gr\"uner}},\ }\href@noop {} {\bibfield  {journal} {\bibinfo  {journal} {Rev.
  Mod. Phys.}\ }\textbf {\bibinfo {volume} {60}},\ \bibinfo {pages} {1129}
  (\bibinfo {year} {1988})}\BibitemShut {NoStop}%
\bibitem [{\citenamefont {Iye}\ \emph {et~al.}(1982)\citenamefont {Iye},
  \citenamefont {Tedrow}, \citenamefont {Timp}, \citenamefont {Shayegan},
  \citenamefont {Dresselhaus}, \citenamefont {Dresselhaus}, \citenamefont
  {Furukawa},\ and\ \citenamefont {Tanuma}}]{PhysRevB.25.5478}%
  \BibitemOpen
  \bibfield  {author} {\bibinfo {author} {\bibfnamefont {Y.}~\bibnamefont
  {Iye}}, \bibinfo {author} {\bibfnamefont {P.~M.}\ \bibnamefont {Tedrow}},
  \bibinfo {author} {\bibfnamefont {G.}~\bibnamefont {Timp}}, \bibinfo {author}
  {\bibfnamefont {M.}~\bibnamefont {Shayegan}}, \bibinfo {author}
  {\bibfnamefont {M.~S.}\ \bibnamefont {Dresselhaus}}, \bibinfo {author}
  {\bibfnamefont {G.}~\bibnamefont {Dresselhaus}}, \bibinfo {author}
  {\bibfnamefont {A.}~\bibnamefont {Furukawa}}, \ and\ \bibinfo {author}
  {\bibfnamefont {S.}~\bibnamefont {Tanuma}},\ }\href@noop {} {\bibfield
  {journal} {\bibinfo  {journal} {Phys. Rev. B}\ }\textbf {\bibinfo {volume}
  {25}},\ \bibinfo {pages} {5478} (\bibinfo {year} {1982})}\BibitemShut
  {NoStop}%
\bibitem [{\citenamefont {Yaguchi}\ \emph {et~al.}(1993)\citenamefont
  {Yaguchi}, \citenamefont {Iye}, \citenamefont {Takamasu},\ and\ \citenamefont
  {Miura}}]{PhysicaB.184.332}%
  \BibitemOpen
  \bibfield  {author} {\bibinfo {author} {\bibfnamefont {H.}~\bibnamefont
  {Yaguchi}}, \bibinfo {author} {\bibfnamefont {Y.}~\bibnamefont {Iye}},
  \bibinfo {author} {\bibfnamefont {T.}~\bibnamefont {Takamasu}}, \ and\
  \bibinfo {author} {\bibfnamefont {N.}~\bibnamefont {Miura}},\ }\href@noop {}
  {\bibfield  {journal} {\bibinfo  {journal} {Physica B: Condensed Matter}\
  }\textbf {\bibinfo {volume} {184}},\ \bibinfo {pages} {332 } (\bibinfo {year}
  {1993})}\BibitemShut {NoStop}%
\end{thebibliography}%

\clearpage
\appendix
\section{Effect of the substrate on the transition}
\begin{figure}
\includegraphics[width=8cm]{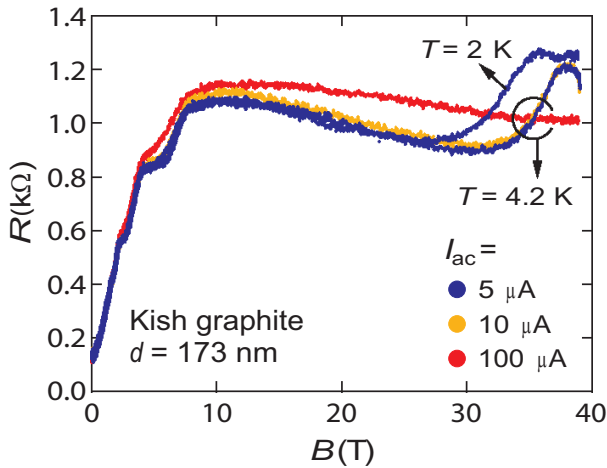}
\caption{In-plane resistance as a function of magnetic field along the $c$ axis in 173-nm-thick graphite for several conditions. At $T=4.2$ K, the semimetal-insulator transition can be observed for low current $5$ and $10\;\mu$A, while larger current of $100\;\mu$A destroys the transition. At $T=2$ K with low current of $5\;\mu$A, the transition shifts to the lower magnetic field. Note that the absolute values of the magnetic fields are not reliable (see main text).}
\label{fA}
\end{figure}

In order to confirm no additional effect on the transition coming from the substrate,
we performed a measurement of the relatively thick sample with $d = 173$ nm.
Figure \ref{fA}
shows the magnetic field dependence of the in-plane resistance for $I_{\textrm{ac}} = 5, 10,$ and $100\;\mu$A at $T = 4.2$ K, in addition to that for $I_{\textrm{ac}} = 5\;\mu$A at $T = 2$ K.
(These data were obtained by a commercial analog lock-in amplification. Owing to the long time constant of the low-pass filter, the absolute value of the magnetic field is not reliable, and the magnetic field dependence of the resistance looks dull.)
In this sample, the transition is clearly observed around $B \simeq 30$ T at $I_{\textrm{ac}} =5$ and $10\;\mu$A,
while it disappears at a higher $I_{\textrm{ac}}$ of 100 $\mu$A.
In a previous report, 
a high dc electric current suppresses the jump of resistance at the transition~\cite{PhysRevLett.54.1182}.
In our measurements,
a higher ac electric current breaks the transition, which seems similar to the dc results,
although it is difficult to distinguish from the Joule heating effect.
In this article, we showed the results obtained by small-enough (dependent on the sample size) ac electric current.
Another characteristic in Fig.~\ref{fA} is that the transition shifts to a lower magnetic field when cooling to $T = 2$ K.
These two features are the same with what was observed in bulk samples~\cite{PhysRevB.25.5478,SolidStateCommun.52.975,PhysicaB.184.332}.
Hence,
the origin of the anomaly in resistance must be the same with that in the bulk,
and we safely conclude that there is no additional effect from the substrate in this kind of sample.

\section{Numerical lock-in technique}
\label{AppendixNumLockin}

\begin{figure}
\centering
\includegraphics[width=8cm]{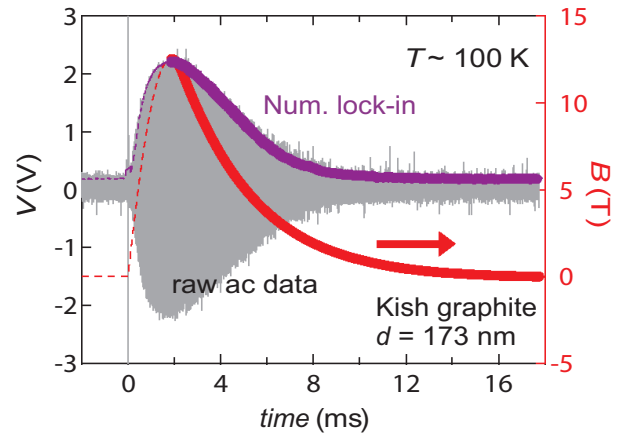}
\caption{Raw ac data in graphite with $d=173$ nm at $T\sim 100$ K, and the dc component extracted from it by the numerical lock-in technique. Magnetic fields are displayed on the right axis. Ascending and descending branches are indicated by broken and solid lines, respectively.}
\label{fB}
\end{figure}

In-plane resistances were obtained by processing ac data with the numerical lock-in technique.
Figure \ref{fB} shows raw ac data and the dc component extracted from it by the numerical lock-in method.
The raw ac data were obtained by a difference amplifier to pick up ac voltage induced by a small ac electrical current.
They were recorded by a fast oscilloscope (1MSa/s).
By multiplying two components of a sinusoidal signal and subsequently filtering with a digital low-pass filter in a numerical manner,
we deduce the amplitude of the dc component.
As mentioned in the main text,
the ascending branch of $\approx 2$ ms is too short for transport measurement,
although it is roughly the same as the descending branch.
In addition, the switching noise, as can be seen at 0 ms, is significant.
This is why we display only the descending branch for the resistance.
If we use a long time constant in the low-pass filter,
it removes not only noise but also the fine structure of the data.
In addition, the dc component extracted from the ac signal will not be synchronized with other dc signals.
Because the magnetic field is measured by a pick-up coil (dc signal),
this is significant for determining an accurate critical field.
We set the time constant to be as short as possible.
Moreover, in order to minimize the difference between dc and ac data,
we modulate the dc data [i.e. $B(time)\times \cos (2\pi f \cdot time)$], where $f = 25$ kHz is the same frequency as that of the resistance measurement,
and perform exactly the same numerical-lock-in process as ac data.
The resultant data, exhibited in Fig.~\ref{f2}, are finally obtained by moving-window averaging.

\section{Density-response function $\chi^{(0)} (q_z)$ in graphite under high magnetic fields}
\label{Appendixchi0calc}

\begin{figure*}
\centering
\includegraphics[width=17cm]{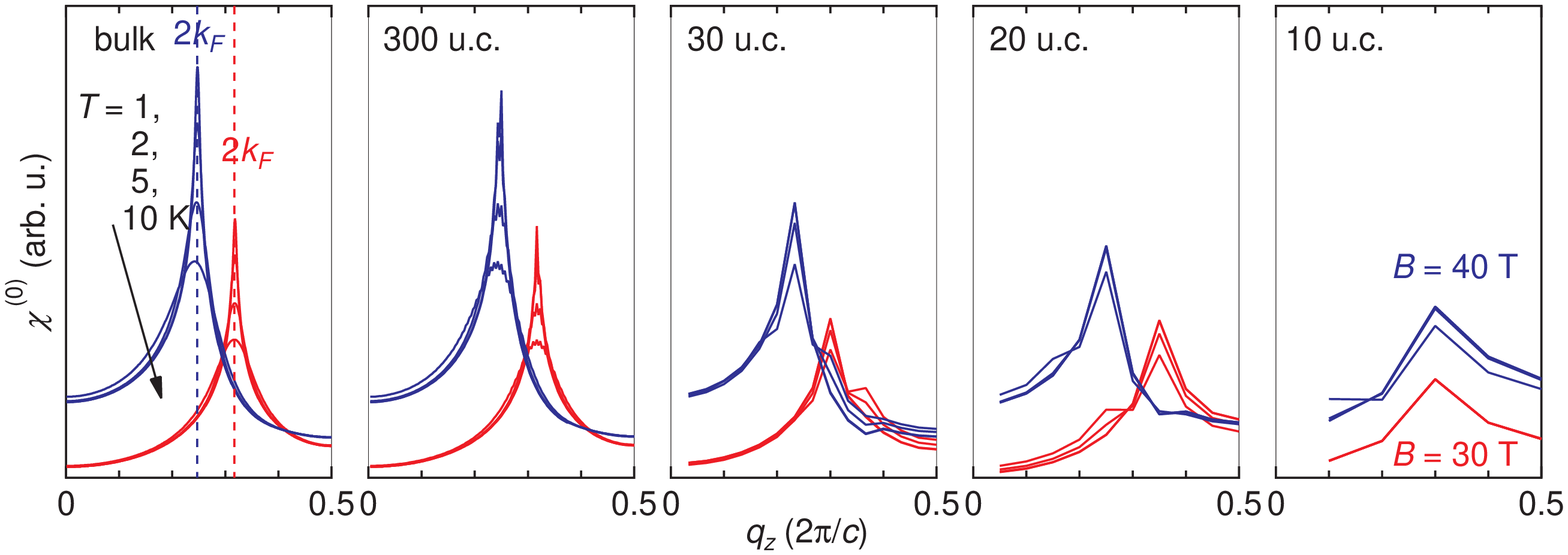}
\caption{Density-response function $\chi^{(0)}$ as a function of $q_z$ under $B = 30$ and 40 T perpendicular to the plate for bulk and 300-, 30-, 20-, and 10-u.c.-thick systems at $T=1,2,5$, and 10 K from top to bottom. For the bulk system, a sharp peak evolves at $q_z = 2k_F$ and is rapidly suppressed with elevating temperatures. These peak-heights shrink with loss of temperature dependence in thin-film systems.}
\label{fC}
\end{figure*}

The simulation of the phase diagram depicted in Fig.~\ref{f3}(b) is provided by an evaluation of the density-response function $\chi^{(0)} (q_z)$, as displayed in Eq.~(\ref{eq:chi0}).
Although some other possible channels exist,
we focus on the ($n=0, \uparrow$) Landau subband for simplicity.
The band dispersion is calculated based on the Slonczewski-Weiss-McClure model for each magnetic field.
In a thin-film system,
the dispersion cannot neglect the discreetness,
so it is natural to evaluate the Fermi energy at each thickness and each magnetic field.
However, in order not to lose the essence,
we simply fix it by the bulk value.
The band dispersions for the bulk and thin-film systems are exemplified in Fig.~\ref{f1}(b).
With these band dispersions,
$\chi^{(0)} (q_z)$ is numerically estimated.
Figure \ref{fC} shows examples of $B = 30$ and 40 T at several temperatures for 300, 30, 20, 10-u.c., and bulk systems.
In bulk and thick-enough (300-u.c.) systems,
$\chi^{(0)} (q_z)$ has a peak structure at $q=2k_F$, and it is rapidly suppressed with elevating temperatures.
On the other hand,
these peak heights shrink in thin-film systems (30, 20, and 10-u.c. thick),
and the temperature dependence becomes moderate.
These features result in the two characteristics in the phase diagram, as discussed in the main text.
Note that the peak position deviates from $2k_F$ in thin-film systems owing to the discreteness of $q_z$.
As discussed in the main text,
the critical condition is that $\max\left[\chi^{(0)}(T,B)\right]$ reaches some critical value $1/\tilde{u}$.
In this simulation,
we set the common condition as $1/\tilde{u}\propto \sqrt{B}$ for all $N$ systems to reproduce the bulk phase diagram.

\end{document}